# Scattering phases for fermion-fermion scattering in the Gross-Neveu model[*]

Meinulf Göckeler[a,b,†], Hans A. Kastrup[b], Jörg Viola[b,a,c], Jörg Westphalen[b,a]

[a]Höchstleistungsrechenzentrum HLRZ, c/o Forschungszentrum Jülich, D-52425 Jülich, Germany

[b]Institut für Theoretische Physik E, RWTH Aachen, D-52056 Aachen, Germany

[c]present address: ZIAM, Kaiserstraße 100, D-52134 Herzogenrath, Germany

We extract scattering phases for fermion-fermion scattering from Monte Carlo simulations of the two-dimensional Gross-Neveu model. This is done by means of Lüscher's method, which exploits the volume dependence of the energies of two-particle states. The results are compared with the analytical predictions.

## 1. INTRODUCTION

Calculating scattering phase shifts nonperturbatively is an important problem, especially in QCD. Lüscher has developed a method to extract elastic scattering phases in massive quantum field theories from two-particle energy levels in a finite volume [1]. Since the latter are accessible to Monte Carlo calculations, it becomes possible to compute scattering phases from numerical simulations. The procedure has been used successfully to study boson-boson scattering in several scalar theories (see e.g. [2] and references therein). Here we report on the first application to a fermionic model: We study fermion-fermion scattering in the two-dimensional Gross-Neveu model.

## 2. THE GROSS-NEVEU MODEL

The action of the Gross-Neveu model for $n_f$ flavours [3] reads in the Euclidean continuum

$$S_c = \int d^2 x \left( \bar{\psi}_a \gamma_\mu \partial_\mu \psi_a - \frac{g^2}{2} (\bar{\psi}_a \psi_a)^2 \right) \quad (1)$$

where $a = 1, 2, \ldots, n_f$. The theory enjoys an $O(2n_f)$ symmetry encompassing the obvious $U(n_f)$ symmetry and charge conjugation. Furthermore it possesses a discrete chiral invariance, which will be called $\gamma_5$ in the following. Since this invariance is spontaneously broken, the model describes $n_f$ massive fermions. The one-particle states (fermions and antifermions) transform according to the vector representation of $O(2n_f)$, whereas the two-particle states are classified according to the representations of $O(2n_f)$ on tensors of rank two: the trivial representation on invariant tensors (abbreviated as inv. in the following), the representations on traceless symmetric tensors (sym.) and on antisymmetric tensors (anti.). The phase shifts in these three symmetry sectors are analytically known [4,5].

On the lattice we use $N$ copies of staggered fermions with "hypercubic" interaction [6]. Hence the action reads

$$\begin{aligned} S = & \sum_{x,\mu} \tfrac{1}{2} \eta_\mu(x) \left( \bar{\chi}_\alpha(x) \chi_\alpha(x+\hat{\mu}) \right. \\ & \left. - \bar{\chi}_\alpha(x+\hat{\mu}) \chi_\alpha(x) \right) \\ & - \frac{g^2}{32} \sum_x \left( \sum_\rho \bar{\chi}_\alpha(x+\rho) \chi_\alpha(x+\rho) \right)^2 \end{aligned} \quad (2)$$

where $\alpha = 1, 2, \ldots, N$ and $\eta_\mu(x)$ are the usual sign factors for staggered fermions. The sum over $\rho$ runs over the corners of an elementary plaquette. For the simulation we rewrite the interaction in a bilinear form by introducing an auxiliary scalar field. Then the hybrid Monte Carlo algorithm is applied. We work on $L \times T$ lattices with the spatial extent $L$ and the temporal extent $T$ varying between 8 and 64.

[*]supported by the *Deutsche Forschungsgemeinschaft*
[†]Speaker at the conference

The discrete chiral invariance of the lattice action is spontaneously broken, and in the continuum limit we obtain $n_f = 2N$ massive fermions. Again, the continuous flavour symmetry is larger than what is immediately apparent. The obvious U($N$) is embedded in the group of unitary symplectic transformations in $2N$ dimensions, USp($N$), which in turn becomes a subgroup of the continuum flavour symmetry O($2n_f$)=O($4N$) in the classical continuum limit [6].

Energy eigenstates with vanishing total momentum, which constitute the starting point of Lüscher's procedure, are classified according to irreducible representations of the group of symmetry transformations leaving the time slices $x_2 =$ const. fixed. The lattice time slice group LTS is generated by the USp($N$) transformations, the spatial shift, the spatial inversion, and the discrete chiral transformation. The continuum time slice group CTS is generated by the O($4N$) transformations, parity, and $\gamma_5$. Note that we can ignore translations as we are only concerned with states of vanishing total momentum.

The irreducible representations of LTS relevant for this work will be denoted by $\Delta_D^{\sigma_1 \sigma_I \sigma_\epsilon}$. The signs $\sigma_1, \sigma_I, \sigma_\epsilon$ correspond to the spatial shift, the spatial inversion, and the discrete chiral transformation, respectively; $D$ denotes an irreducible representation of USp($N$) on tensors of rank two and takes the values inv. (trivial representation on invariant tensors), anti. (representation on antisymmetric tensors with vanishing symplectic trace), and sym. (representation on symmetric tensors). Similarly we have irreducible representations $\bar\Delta_{\bar D}^{\sigma_P \sigma_5}$ of CTS with signs $\sigma_P, \sigma_5$ corresponding to parity and $\gamma_5$, respectively, and $\bar D$ denoting an irreducible representation of O($4N$) on tensors of rank two.

Restriction of an irreducible CTS representation to the subgroup LTS will in general lead to a reducible representation. It turns out that every $\Delta_D^{\sigma_1 \sigma_I \sigma_\epsilon}$ couples to two continuum symmetry sectors $\bar\Delta_{\bar D}^{\sigma_P \sigma_5}$ with different values of $\sigma_P$ (see Table 1). This is analogous to the appearance of the parity partners in QCD with staggered fermions. Whenever a given continuum symmetry sector can be reached by lattice operators from several

Table 1: Connection between the irreps of CTS and LTS. "Suppressed" contributions are shown in italics, "preferred" contributions are underlined.

| LTS $\Delta_D^{\sigma_1 \sigma_I \sigma_\epsilon}$ | | | | CTS $\bar\Delta_{\bar D}^{\sigma_P \sigma_5}$ | | | | | |
|---|---|---|---|---|---|---|---|---|---|
| $D$ | $\sigma_1$ | $\sigma_I$ | $\sigma_\epsilon$ | $\bar D$ | $\sigma_P$ | $\sigma_5$ | $\bar D$ | $\sigma_P$ | $\sigma_5$ |
| inv. | − | − | − | <u>anti.</u> | − | + | <u>anti.</u> | + | − |
| inv. | − | − | + | <u>anti.</u> | − | − | <u>anti.</u> | + | + |
| inv. | − | + | − | *anti.* | + | + | <u>anti.</u> | − | − |
| inv. | − | + | + | *anti.* | + | − | <u>anti.</u> | − | + |
| inv. | + | − | − | *inv.* | − | − | <u>anti.</u> | + | + |
| inv. | + | − | + | *inv.* | − | + | <u>anti.</u> | + | − |
| inv. | + | + | − | <u>inv.</u> | + | − | anti. | − | + |
| inv. | + | + | + | <u>inv.</u> | + | + | <u>anti.</u> | − | − |
| anti. | − | − | − | <u>anti.</u> | + | − | <u>anti.</u> | − | + |
| anti. | − | − | + | <u>anti.</u> | + | + | <u>anti.</u> | − | − |
| anti. | − | + | − | <u>anti.</u> | − | − | *anti.* | + | + |
| anti. | − | + | + | <u>anti.</u> | − | + | *anti.* | + | − |
| anti. | + | − | − | *sym.* | − | − | <u>anti.</u> | + | + |
| anti. | + | − | + | *sym.* | − | + | *anti.* | + | − |
| anti. | + | + | − | <u>sym.</u> | + | − | anti. | − | + |
| anti. | + | + | + | <u>sym.</u> | + | + | <u>anti.</u> | − | − |
| sym. | − | − | − | <u>sym.</u> | + | − | sym. | − | + |
| sym. | − | − | + | <u>sym.</u> | + | + | sym. | − | − |
| sym. | − | + | − | *sym.* | − | − | <u>sym.</u> | + | + |
| sym. | − | + | + | *sym.* | − | + | <u>sym.</u> | + | − |
| sym. | + | − | − | *anti.* | − | − | <u>sym.</u> | + | + |
| sym. | + | − | + | anti. | − | + | <u>sym.</u> | + | − |
| sym. | + | + | − | *anti.* | + | − | *sym.* | − | + |
| sym. | + | + | + | *anti.* | + | + | *sym.* | − | − |

lattice symmetry sectors one has the possibility to check the influence of lattice artifacts.

## 3. TWO-PARTICLE ENERGIES

The needed two-particle energies are extracted from correlation functions

$$C_{ij}(\tau) = \langle \mathcal{O}_i(\tau) \mathcal{O}_j(0) \rangle \,, \; i,j = 1, 2, \ldots, r, \qquad (3)$$

where the operators $\mathcal{O}_i$ have the following properties:

- $\mathcal{O}_i(\tau)$ lives on the two adjacent time slices $x_2 = \tau, \tau + 1$;

- $\mathcal{O}_i(\tau)$ is bilinear in the fermion fields and besides $\bar\chi\chi$ terms it may also contain terms of the form $\chi\chi$ and $\bar\chi\bar\chi$;

- $\mathcal{O}_i(\tau)$ has vanishing total momentum;

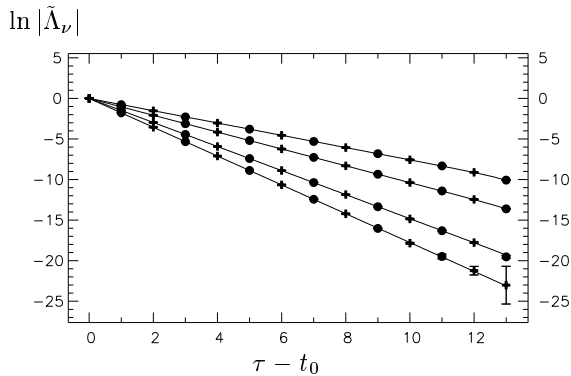

Figure 1. Generalized eigenvalues $\tilde{\Lambda}_\nu$ for the LTS irrep $\Delta^{+--}_{\text{sym.}}$ on a $16 \times 32$ lattice. The crosses (dots) correspond to positive (negative) values of $\tilde{\Lambda}_\nu$.

- $\mathcal{O}_1(\tau), \ldots, \mathcal{O}_r(\tau)$ belong to one definite irreducible representation $\Delta^{\sigma_1 \sigma_I \sigma_\epsilon}_D$ of LTS.

From the transfer matrix formalism we obtain the following representation of the correlation functions $C_{ij}(\tau)$ as $T \to \infty$:

$$C_{ij}(\tau) = \sum_n A^{(n)}_{ij} \left(-\sigma_I \sigma_P^{(n)}\right)^\tau \\ \times \left(e^{-E_n \tau} + e^{-E_n(T-\tau)}\right). \quad (4)$$

Therefore the continuum quantum number $\sigma_P^{(n)}$ and hence the complete continuum symmetry sector to which a particular two-particle energy level $E_n$ belongs can be read off from the oscillating or non-oscillating character of the corresponding contribution to the correlation function and the lattice quantum numbers of the operators used ($\sigma_I$, in particular).

However, in the data (for $n_f = 8$) either all contributions are oscillating or they are all non-oscillating or one hardly sees any signal. How can this behaviour be explained? First, one observes that certain combinations of continuum quantum numbers cannot be realized by two-particle states due to the Pauli principle: For $\bar{D}$ = anti. contributions with $\sigma_P = +1$ are suppressed as is $\sigma_P = -1$ for $\bar{D}$ = inv., sym. Secondly, in the continuum limit our operators, which are nonlocal in time, will couple predominantly to one value of $\sigma_P$, an effect that was also observed for nonlocal-in-time meson operators in QCD with staggered fermions. For $\bar{D}$ = anti., inv. contributions with $\sigma_P = -\sigma_\epsilon \sigma_1^{(1-\sigma_\epsilon)/2}$ are preferred, whereas for $\bar{D}$ = sym. the preferred value is $\sigma_P = \sigma_\epsilon \sigma_1^{(1-\sigma_\epsilon)/2}$. In the end it turns out that each lattice operator couples (effectively) to two-particle states from at most one continuum symmetry sector (see Table 1).

We are now ready to discuss the computation of the two-particle energies. Since in practice $T$ is not extremely large, there are additional contributions of the form $(-1)^\tau a_{ij} + b_{ij}$ in $C_{ij}(\tau)$. They should be of the order $e^{-m_f T}$ and result from elementary fermions of mass $m_f$ travelling once around the lattice in time direction. We eliminate them by subtracting $C_{ij}(\tau + 2)$. Exploiting the symmetry in $\tau$ we arrive at the $r \times r$ correlation function matrix

$$\begin{aligned}\tilde{C}_{ij}(\tau) &= C_{ij}(\tau) - C_{ij}(\tau + 2) \\ &\quad + C_{ij}(T - \tau) - C_{ij}(T - \tau - 2) \\ &= \sum_n A^{(n)}_{ij} \Lambda(\tau, E_n) \end{aligned} \quad (5)$$

with

$$\begin{aligned}\Lambda(\tau, E_n) &= 8 e^{-E_n T/2} \left(-\sigma_I \sigma_P^{(n)}\right)^\tau \sinh(-E_n) \\ &\quad \times \sinh\left(E_n(\tau + 1 - \tfrac{1}{2} T)\right). \end{aligned} \quad (6)$$

Next we solve the generalized eigenvalue problem [7]

$$\sum_{j=1}^r \tilde{C}_{ij}(\tau) w^\nu_j = \frac{\Lambda(\tau, E_\nu)}{\Lambda(t_0, E_\nu)} \sum_{j=1}^r \tilde{C}_{ij}(t_0) w^\nu_j \quad (7)$$

($\nu = 1, 2, \ldots, r$) for fixed $t_0$ ($t_0 = 1$ in the following). From the corresponding eigenvalues $\tilde{\Lambda}_\nu(\tau) \equiv \Lambda(\tau, E_\nu)/\Lambda(t_0, E_\nu)$ the energies $E_1 < E_2 < \ldots < E_r$ are then obtained by a one-parameter fit. (The influence of the neglected states above $E_r$ can be tested by varying $r$, the number of operators considered.)

In Fig. 1 we plot $\ln |\tilde{\Lambda}_\nu|$ versus $\tau - t_0$ for one lattice symmetry sector. Positive values of $\tilde{\Lambda}_\nu$ are indicated by plus signs, negative values by dots. The curves result from one-parameter fits. The number of continuum flavours is 8, and the coupling is such that $m_f = 0.401(2)$.

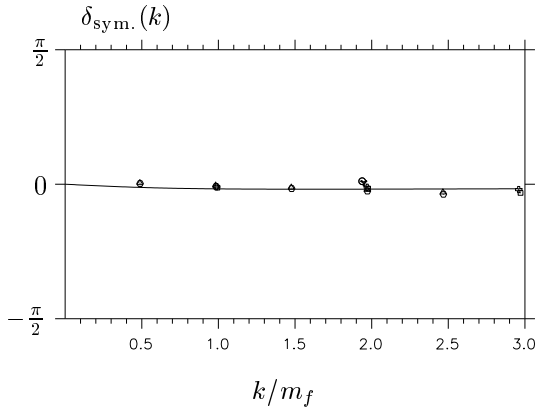

Figure 2. Scattering phase from the LTS irreps $\Delta^{+-+}_{\text{sym.}}$ and $\Delta^{++-}_{\text{anti.}}$ (which all couple to the sym. sector in the continuum limit, see Table 1) for lattices $8 \times 8$, $16 \times 32$, $32 \times 32$.

## 4. THE SCATTERING PHASE SHIFTS

According to Lüscher, the scattering phase $\delta$ at the momentum $k_\nu$ related to the two-particle energy $E_\nu$ through

$$\tfrac{1}{2} E_\nu = \sqrt{m_f^2 + k_\nu^2} \tag{8}$$

is given by

$$\delta(k_\nu) = -\frac{L}{2} k_\nu \mod \pi \tag{9}$$

(provided polarization effects are negligible) [1,7]. So here the continuum dispersion relation enters. Since in the fermion propagator the lattice dispersion relation for free fermions turns out to be very well satisfied we feel encouraged to replace the continuum dispersion relation by the lattice dispersion relation in an attempt to compensate cutoff effects at least partially [2]. Hence we determine $k_\nu$ from

$$\sinh(\tfrac{1}{2} E_\nu) = \sqrt{m_f^2 + \sin^2 k_\nu} \,. \tag{10}$$

The scattering phase $\delta$ at $k_\nu$ is then taken from Eq. (9). In Figs. 2,3 we show the results for two continuum symmetry sectors coming from a variety of lattice symmetry sectors and lattice sizes. Thus the plots give an idea of the systematic errors involved.

The curves represent the analytically calculated phase shifts [4,5]. We observe good agreement up to $k/m_f \approx 3$ showing that Lüscher's method works also for fermions – at least in two dimensions. The deviations for larger momenta may be due the fact that the higher energy levels do no longer correspond to pure two-particle states.

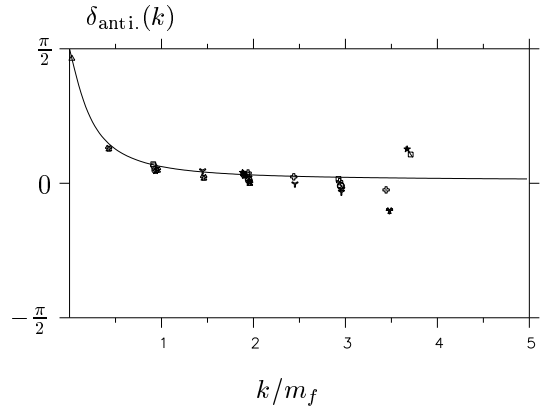

Figure 3. Scattering phase from the LTS irreps $\Delta^{-+-}_{\text{anti.}}$, $\Delta^{--+}_{\text{anti.}}$, $\Delta^{+++}_{\text{anti.}}$ and $\Delta^{+--}_{\text{sym.}}$ (which all couple to the anti. sector in the continuum limit, see Table 1) for lattices $8 \times 8$, $16 \times 32$, $32 \times 32$.

## ACKNOWLEDGEMENT

We wish to thank the HLRZ Jülich for providing the necessary computer time.